\documentclass[letter,twocolumn]{jpsj3}
\usepackage{txfonts}
\usepackage[usenames]{color}
\usepackage{graphicx}
\usepackage{dcolumn}
\usepackage{bm}

\def\gsim{\buildrel {\textstyle >}\over {_\sim}}
\def\lsim{\buildrel {\textstyle <}\over {_\sim}}

\title{
Slow-Slip Phenomena Represented by the One-Dimensional Burridge-Knopoff Model of Earthquakes
}

\author{
Hikaru Kawamura\thanks{E-mail:kawamura@ess.sci.osaka-u.ac.jp}, Maho Yamamoto and Yushi Ueda
}

\inst{                    
Department of Earth and Space Science, Faculty of Science,
Osaka University, \\
Toyonaka 560-0043, Japan
}

\abst{
Slow-slip phenomena, including afterslips and silent earthquakes, are studied using a one-dimensional Burridge--Knopoff model that obeys the rate-and-state dependent friction law. By varying only a few model parameters, this simple model allows reproducing a variety of seismic slips within a single framework, including main shocks, precursory nucleation processes, afterslips, and silent earthquakes.
}

\begin{document}
\maketitle

An earthquake is a stick-slip dynamic instability of a pre-existing fault driven by the motion of a tectonic plate \cite{Scholz2002}. Numerical simulations of earthquakes based on a simplified statistical model, the so-called Burridge--Knopoff (BK) model \cite{BK}, are popular in statistical physics, and they provide considerable information about the statistical properties of earthquakes \cite{CLS,Kawamura-review}. Although the BK model has been successful in describing earthquakes, almost all studies so far have been limited to the high-speed rupture of earthquakes or to main shocks. 

Meanwhile, recent development in modern GPS technology and in high-density GPS and seismograph networks has revealed a rich variety of slow-slip phenomena, including afterslips \cite{Kawasaki1999,Heki}, silent earthquakes \cite{Hirose,Dragert,Miller,Ozawa,Kostoglodov,Kawasaki2004}, deep tremors \cite{Obara}, {\it etc.\/}, where the fault sliding velocity is several orders of magnitudes slower than that of the standard high-speed rupture. Thus, the concept of seismicity has been broadened dramatically \cite{Beroza}. Then, to gain a complete understanding of earthquake phenomena, one needs to incorporate these slow-slip phenomena.

 It is a challenge to understand such a wide variety of seismicity from a general physical viewpoint, including slow slips. Therefore, questions such as what are the characteristics of slow-slip phenomena, how it differs from the standard high-speed rupture of a main shock, what conditions cause them to occur, {\it etc.\/}, need to be answered. In the present letter, we wish to address this issue from the statistical-physics viewpoint by employing the 1D BK model obeying the rate-and-state dependent friction (RSF) law\cite{Dieterich,Ruina,Scholz1998}. We successfully reproduce a variety of seismic phenomena, including high-speed rupture of main shocks, its precursory nucleation processes, afterslips, and silent earthquakes, by varying only a few fundamental parameters of the model. In particular, regarding the occurrence of slow-slip phenomena, the relative magnitude of the frictional parameters $a$ and $b$ characterizing the RSF law turns out to be crucial. 

 The 1D BK model obeying the RSF law considered here consists of a 1D array of $N$ identical blocks with mass $m$, which are mutually connected by two neighboring blocks via elastic springs with spring stiffness $k_c$, and these are connected to a moving plate via springs with spring stiffness $k_p$. All blocks are subject to a friction force $\Phi$. The dimensionless equation of motion for the $i$-th block can be written as \cite{CLS,Kawamura-review}
\begin{eqnarray}
\frac{d^2u_i}{dt^2} = \nu t-u_i+l^2(u_{i+1}-2u_i+u_{i-1}) - \phi_i ,
\label{eq-motion}
\end{eqnarray}
where $\ell \equiv (k_c/k_p)^{1/2}$. The dimensionless displacement $u_i$ is normalized by the characteristic slip distance ${\mathcal L}$ associated with the RSF law, the time $t$ by $\omega^{-1}=\sqrt{m/k_p}$, the block velocity $v_i$ and the pulling speed of the plate $\nu$ by $\mathcal{L}\omega$, and the dimensionless friction force $\phi$ by $k_p{\mathcal L}$.

 The RSF force $\phi_i$ assumed here reads as \cite{Ueda2014,Ueda2015,Kawamura2017}
\begin{equation}
\phi_i=c+a\log(1+\frac{v_i}{v^*})+b\log \theta_i ,
\label{RSF}
\end{equation}
where $\theta_i$ is the dimensionless state variable describing the ``state'' of the interface, and the normalized friction parameters $a$, $b$, and $c$ represent velocity-strengthening, velocity-weakening, and constant parts of friction, respectively. The original friction parameters $A$, $B$, and $C$ are related to the normalized parameters by $A=(k_p{\mathcal L}/{\mathcal N})a$, $B=(k_p{\mathcal L}/{\mathcal N})b$, and $C=(k_p{\mathcal L}/{\mathcal N})c$, where ${\mathcal N}$ is the normal load. For simplicity, we inhibit the motion in the direction opposite to the motion of the plate. The state variable $\theta_i$ is assumed to obey the aging law \cite{Ruina},
\begin{equation}
\frac{{\rm d}\theta_i}{{\rm d}t} = 1-v_i\theta_i .
\label{aging}
\end{equation}

Note that, in contrast to the standard RSF law with the $a$ term being proportional to a pure logarithmic form $\log v$, which yields a pathological limit of a negatively divergent friction for $v\rightarrow 0$, we phenomenologically introduce a modified form using the crossover velocity $v^*$  \cite{Ueda2014,Ueda2015,Kawamura2017}. The modified form, where the $a$ term reduces to a purely logarithmic form when $v>>v^*$ but becomes proportional to the block velocity $v$ when $v<<v^*$, is able to describe a complete halt, unlike the standard form. 

 The model is characterized by only a few basic dimensionless parameters, {\it e.g.\/}, frictional parameters $a$, $b$, and $v^*$ ($c$ is actually irrelevant \cite{Kawamura2017}), and the elastic parameter $\ell$. The plate velocity $\nu$ is also unimportant so long as it remains small, aside from setting the interseismic timescale. 

 Estimates of typical values of the model units representing natural faults have been given \cite{Ueda2014}. The BK model possesses a built-in time scale $\omega^{-1}$ corresponding to the rise time of an earthquake event. This may be estimated to be $\sim 1$ [s]. The model possesses two distinct and independent length scales: one associated with the fault slip and the other with the distance along the fault. The former length scale is the critical slip distance ${\mathcal L}$, which was estimated to be $\simeq 1$ [cm], while the other is the distance the rupture propagates per unit time, $v_s/\omega$, which was estimated to be $\simeq 2-3$ [km], where $v_s$ is the $s$-wave velocity along the fault. The typical plate velocity is several [cm/year] and corresponds to a very small number of $\nu\sim 10^{-7}-10^{-8}$. The spring constant $k_p$ was related to the normal stress as $\frac{{\mathcal N}}{k_p{\mathcal L}}\simeq 10^2-10^3$. Then, as $C$ is known to take a value around $\frac{2}{3}$, $c$ would be of order $10^2$-$10^3$, with $a$ and $b$ being one or two orders of magnitude smaller than $c$. The crossover velocity $v^*$ is hard to estimate, though it should be much smaller than unity, and we take it as a parameter.

 The properties of the high-speed rupture of the model were investigated,\cite{CaoAki,OhmuraKawamura,Kawamura2017} together with the properties of its precursory nucleation process.\cite{Ueda2014,Ueda2015} The two regimes with mutually different seismic properties are identified \cite{Ueda2014,Ueda2015,Kawamura2017} as the weak and strong frictional instability regimes. In the former, the frictional instability parameter $b$ is smaller than a critical value $b_c=2\ell^2+1$ determined by the elastic parameter $\ell$, and the main shocks possess eminently ``characteristic'' features \cite{Kawamura-review,Kawamura2017} accompanied by a nucleation process with a quasi-static, long-lasting initial phase. In the latter, $b>b_c$, and main shocks possess more or less ``critical'' features\cite{Kawamura-review,Kawamura2017} with a nucleation process that is unaccompanied by the initial phase. These earlier studies on the model concentrated on the parameter regime where the high-speed rupture dominates, which corresponded to $a<b$, {\it i.e.\/}, the frictional weakening dominates over strengthening. In fact, in such a ``main-shock regime'', the parameters $a$ and $v^*$ turned out to be irrelevant.

 Generally, since the frictional parameters $a$ and $b$ compete in their function, either $a<b$ or $a>b$ might significantly affect the model dynamics. More specifically, one may expect that, when $a \gsim b$, the compensation effect due to the $a$ term might induce slow-slip phenomena in the model. In the present letter, we perform a systematic survey of the parameter range of $a \gsim b$, where the velocity strengthening $a$ term is expected to play a significant role. We find that slow slips actually come into play there. When $a$ is comparable to or only a few times larger than $b$, say, $b \lsim a \lsim 2b$, the model still exhibits a main shock of high-speed rupture, but subsequently exhibits a slow and long-lasting afterslip. We call this parameter regime ``region II'' in distinction with ``region I'' of high-speed rupture where $a \lsim b$. When $a$ is considerably larger than $b$, say, $a \gsim 2b$, the model no longer exhibits high-speed rupture, but instead exhibits a long-lasting slow slip only, {\it i.e.\/}, a silent earthquake. We call this parameter regime ``region III''.

 In our simulations, for the sake of computational feasibility, we concentrate on the strong frictional instability regime where $b>b_c$ by setting $\ell=3$ ($b_c=19$) and $b=30(>b_c)$. The parameters $a$ and $v^*$ are varied in the range of $1-240$ and $10^{-1}-10^{-3}$, respectively, while fixing $\nu=10^{-7}$. The system size (total number of blocks) is 800, and open boundary conditions are imposed.  Concerning the initial conditions, all blocks are assumed to be at rest, {\it i.e.\/}, $v_i=0$ ($1\leq i\leq N$) at $t=0$, the state variable is taken to be uniform $\theta_i=10^8$, while the displacement of each block is assumed to take random values uniformly distributed between -0.5 and 0.5 from block to block. Events at earlier times are transient and are affected in a non-stationary manner by the initial conditions. We wait until the system reaches the stationary state and loses its initial memory, and we compute various observables in such stationary states.

\begin{figure}
\includegraphics[scale=0.39]{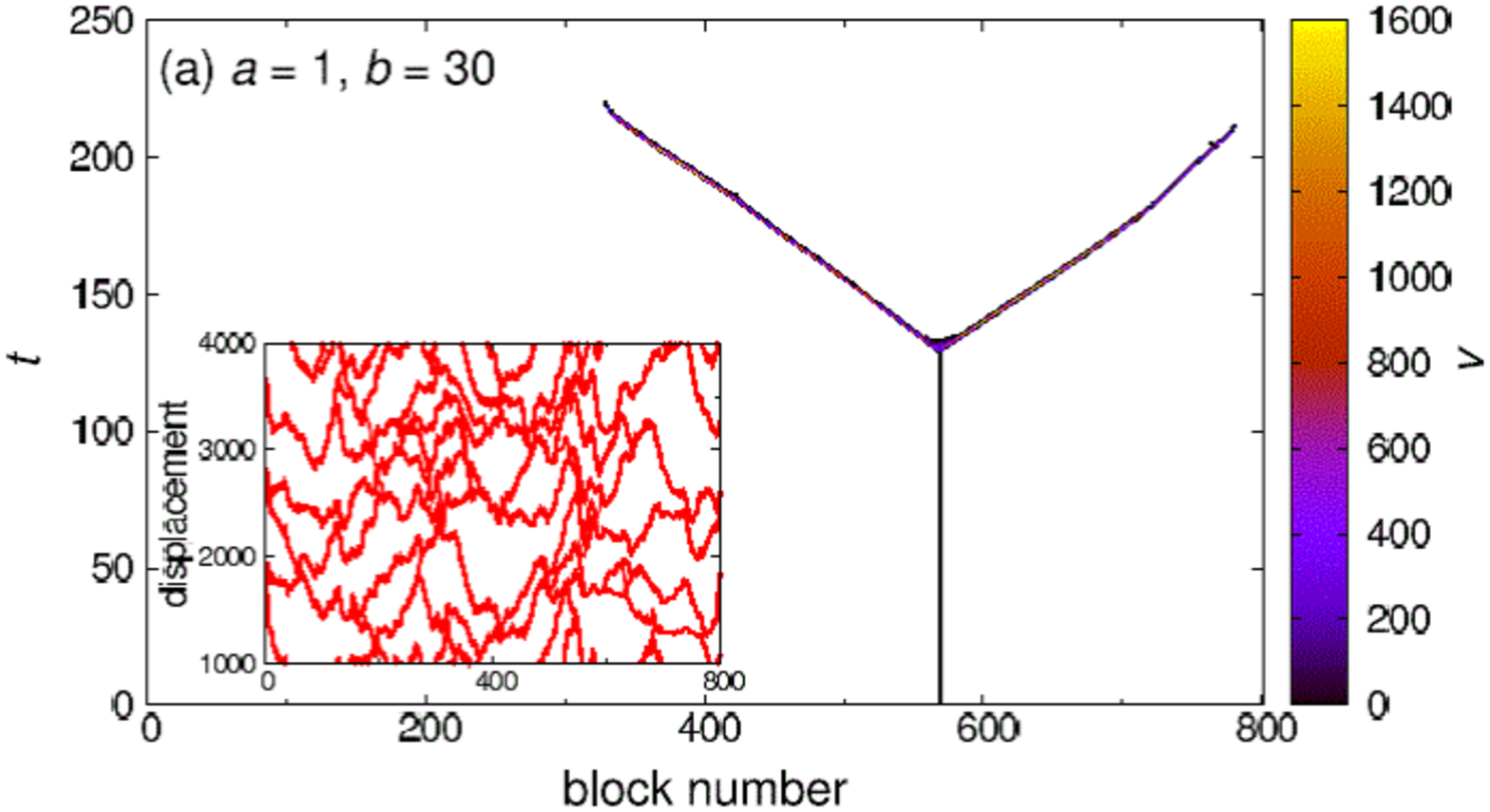}
\includegraphics[scale=0.39]{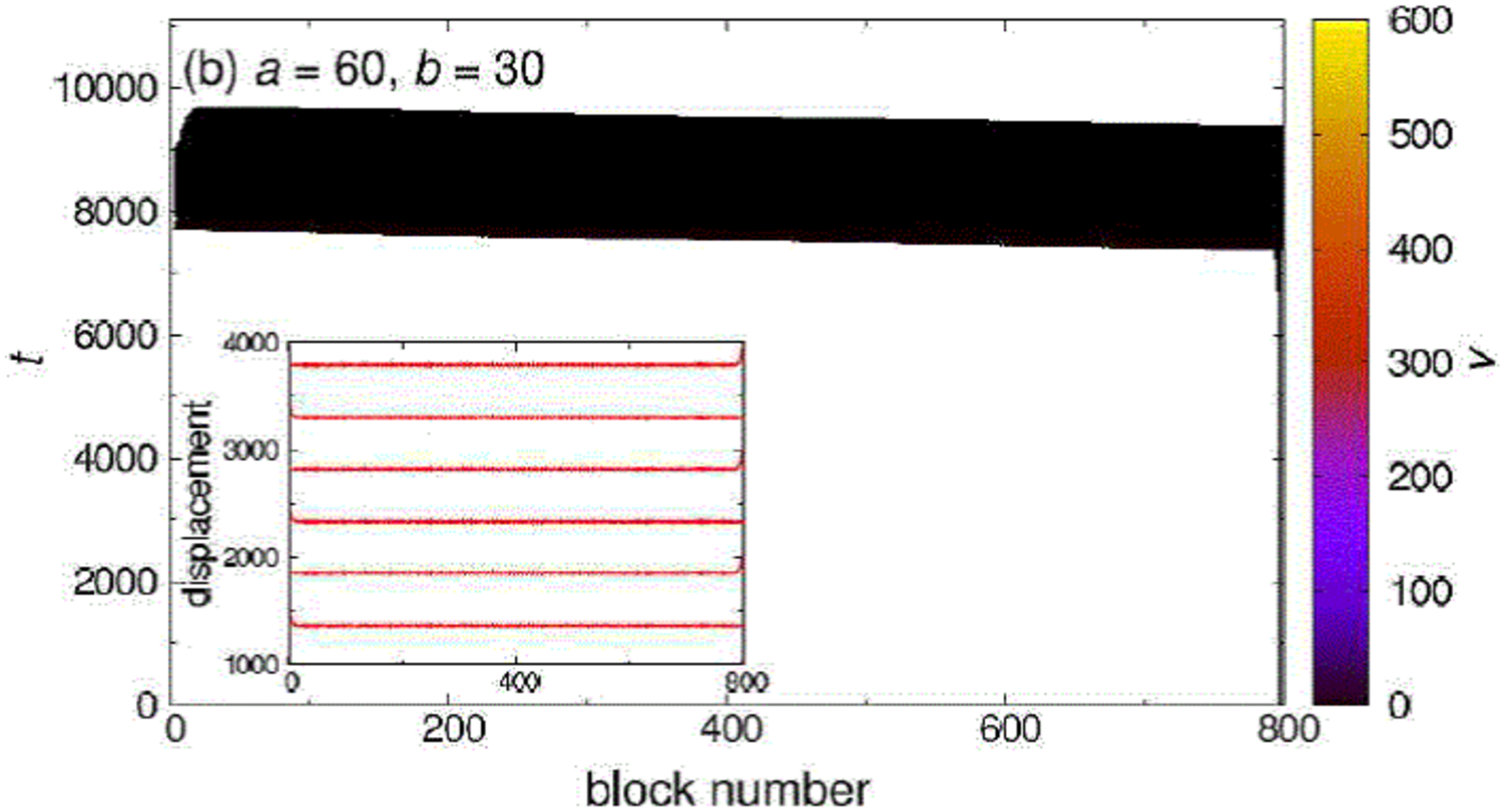}
\includegraphics[scale=0.39]{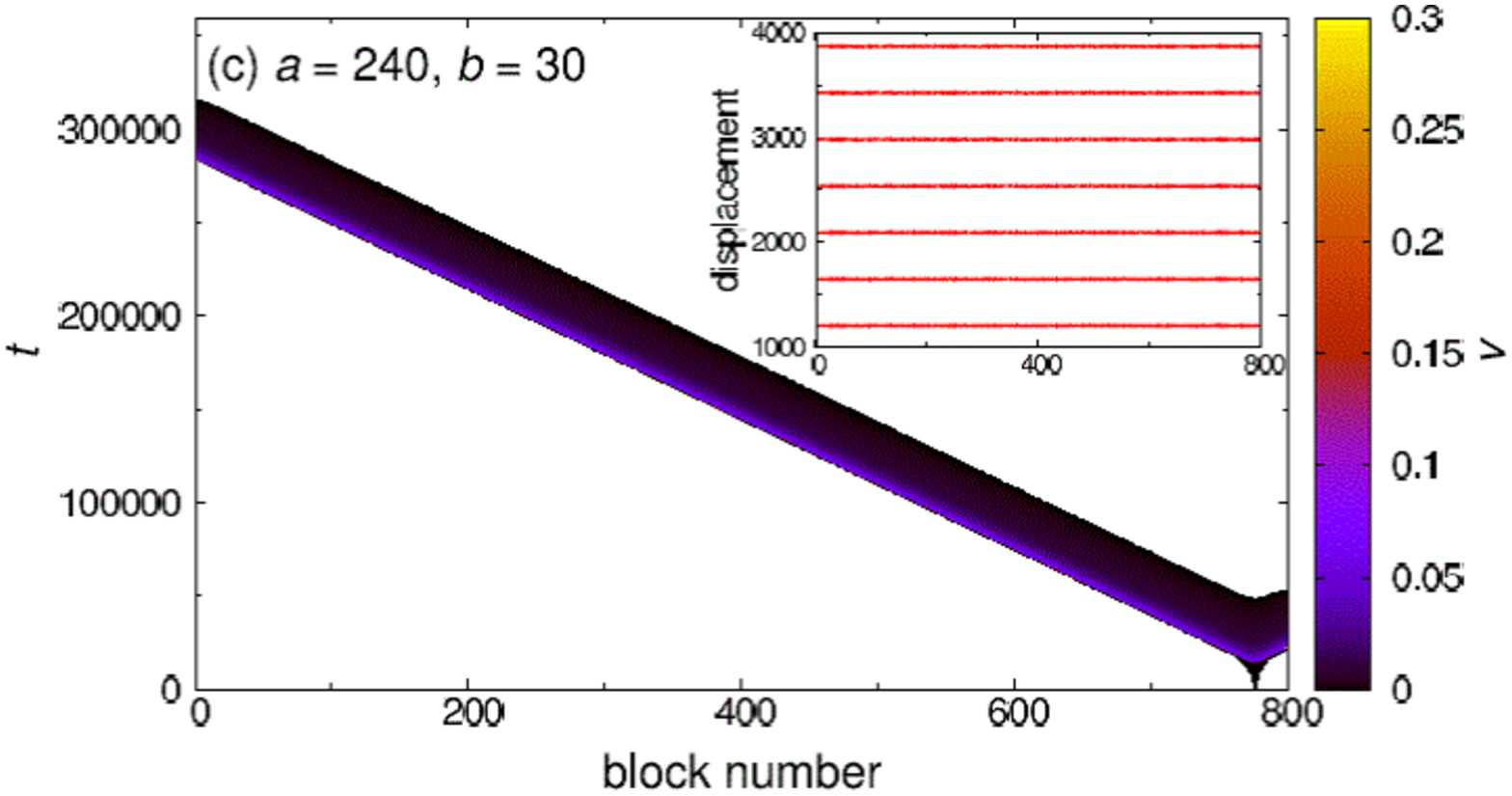}
\caption{
(Color online) Plots of the block sliding velocity on the position (block number) versus time. White indicates a complete halt. The $a$ parameter is (a) $a=1$ (region I), (b) $a=60$ (region II), and (c) $a=240$ (region III). Other parameters are $b=30$, $c=1000$, $v^*=10^{-2}$, $\ell=3$, and $\nu=10^{-7}$. The time origin $t=0$ is taken at the beginning of the nucleation process of the event.  The insets show position versus time plots of the event sequences in the stationary state, where the curves are drawn when all blocks are at rest.
}
\end{figure}

 In Figs.1(a)-(c), we show the space-time evolution of a typical event in the stationary state of the model as color plots of the sliding velocity, where (a) $a=1$ corresponds to region I (high-speed rupture regime), (b) $a=60$ corresponds to region II (aftershock regime), and (c) $a=240$ corresponds to region III (silent earthquake regime), with $v^*=10^{-2}$. The diagonal lines in Figs.(a) and (b) emanating from the nucleation site represent the propagating mode of a main shock, and the ones in Fig.(c) of a silent earthquake. The black ``bands'' in Figs.(b) and (c) represent the long-lasting slow-slip parts, {\it i.e.\/}, an afterslip in (b) and a silent earthquake in (c). Note the big difference in time scale as represented by the units on the vertical axis in each figure.

 For events accompanying slow slips, we show in Fig.2 the time $t$ dependence of the sliding velocity $v$ on a log-log plot, for a typical block located in the rupture-propagating part. The data for $v^*=10^{-3}$ are also shown together with those for $v^*=10^{-2}$. For events belonging to region II, the maximum velocity of $v_{max}\simeq 10^1-10^3$ is reached and is characteristic of a high-speed rupture, which is followed by long-lasting slow afterslips. In contrast, $v_{max}$ of the events belonging to region III is of order $10^{-1}-10^{-2}$, orders of magnitudes smaller than that of a high-speed rupture. Hence, the events in region III might well be identified as silent earthquakes. Its duration time is longer than that of a main shock in region I by a factor of approximately $10^4-10^5$ for $v^*=10^{-3}$. As $v$ gets smaller beyond $v^*$, the block comes to a stop fairly soon. Namely, $v^*$ signals the end of the slow-slip process. Then, a longer-lasting slow slip is naturally expected for a smaller $v^*$.

\begin{figure}
\includegraphics[scale=0.5]{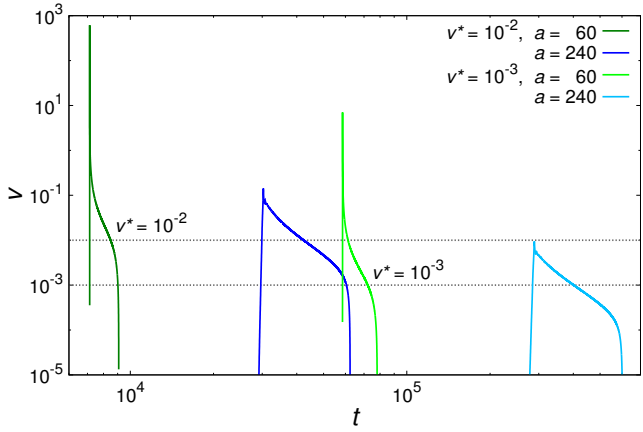}
\caption{
(Color online) The time $t$ dependence of the sliding velocity $v$ for a block located in the rupture-propagating part for typical events belonging to region II ($a=60$, $v^*=10^{-2}$, and $10^{-3}$) and region III ($a=240$, $v^*=10^{-2}$ and $10^{-3}$). The other parameters are $b=30$, $c=1000$, $\ell=3$, and $\nu=10^{-7}$. The time origin $t=0$ is taken at the beginning of the nucleation process of the event.
}
\end{figure}
\begin{figure}
\includegraphics[scale=0.45]{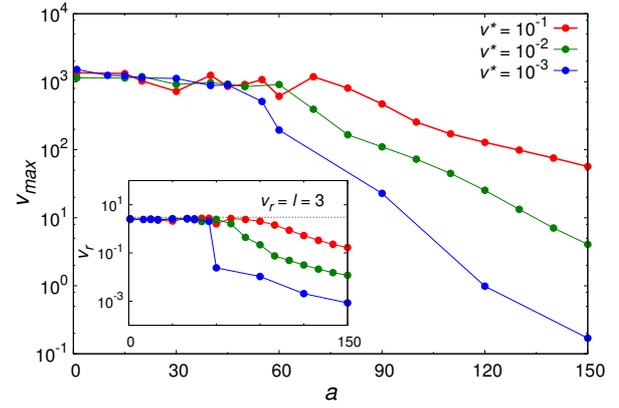}
\caption{
(Color online) The $a$ dependence of the maximum sliding velocity $v_{max}$ of events for $v^*=10^{-1}$, $10^{-2}$, and $10^{-3}$. The other parameters are $b=30$, $c=1000$, $\ell=3$, and $\nu=10^{-7}$. The inset shows the corresponding plot of the rupture-propagation velocity $v_r$.
}
\end{figure}

 One might wonder if the distinction between regions II and III were only gradual. We show in Fig.3 the $a$-dependence of $v_{max}$ spanning both regions II and III for various $v^*$-values. As can be seen from the figure, a clear change of behavior occurs around $a\simeq 50-60$, although the borderline value seems slightly $v^*$-dependent. A similar behavior is observed in the $a$-dependence of the rupture-propagating velocity $v_r$ shown in the inset, which corresponds to the inverse slope of the diagonal lines of Fig.1. We regard the observed change of behavior, which tends to be more eminent for smaller $v^*$, as a signature discriminating regions II and III. Presumably, while the change is a crossover for a finite $v^*$-value, it may become a sharp ``transition'' in the $v^*\rightarrow 0$ limit.

 Regions I and II are discriminated by the absence/presence of an afterslip following the high-speed rupture of a main shock. This change turns out to occur roughly around $a\sim b$. In the insets of Figs.1(a)-(c), we show typical event sequences in the stationary state, in which the displacement of each block is plotted each time all blocks are at rest. As can be seen from Fig.1(a), all events in region I are ``partial'' breaking in particular parts of the entire system with nontrivial size distribution. By contrast, as can be seen from Fig.1(c), events in region III are ``entire'' breaking the entire system that repeat quasi-periodically. The situation in region II is a bit more complex. In the parameter region close to region III, events tend to be ``entire'', as shown in the inset of Fig.1(b). Meanwhile, in the parameter region close to region I, the system is often divided into several large sectors, a situation somewhat intermediate between ``partial'' (region I) and ``entire'' (region III) ruptures.

 In the main shock of region I, the stress distribution after the event tends to be spatially inhomogeneous, which serves as a ``barrier'' for subsequent events. Meanwhile, in the events in regions II and III, slow slips tend to smoothen such spatial inhomogeneity, eliminating the potential barrier for subsequent events.

\begin{figure}
\includegraphics[scale=0.45]{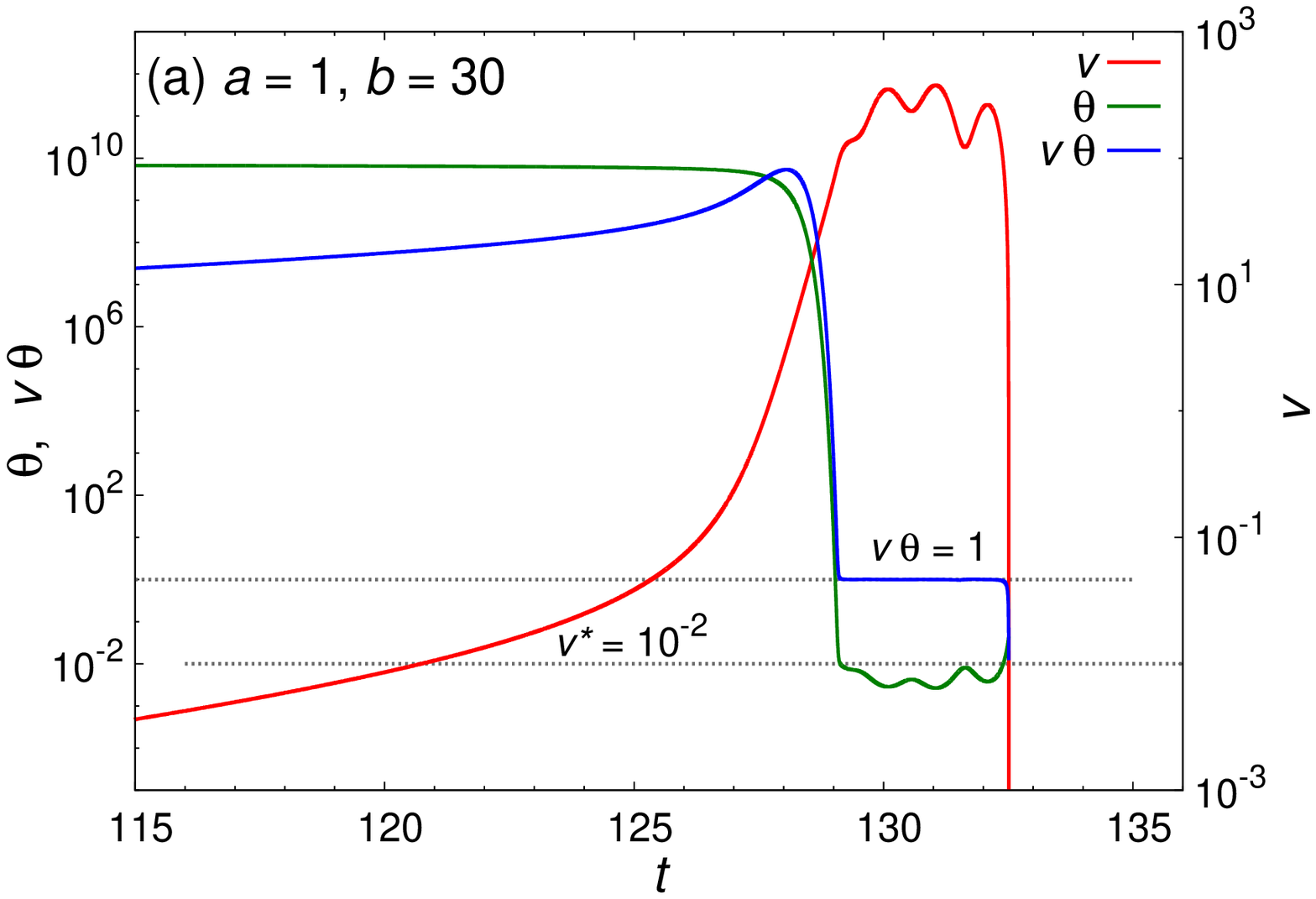}
\includegraphics[scale=0.45]{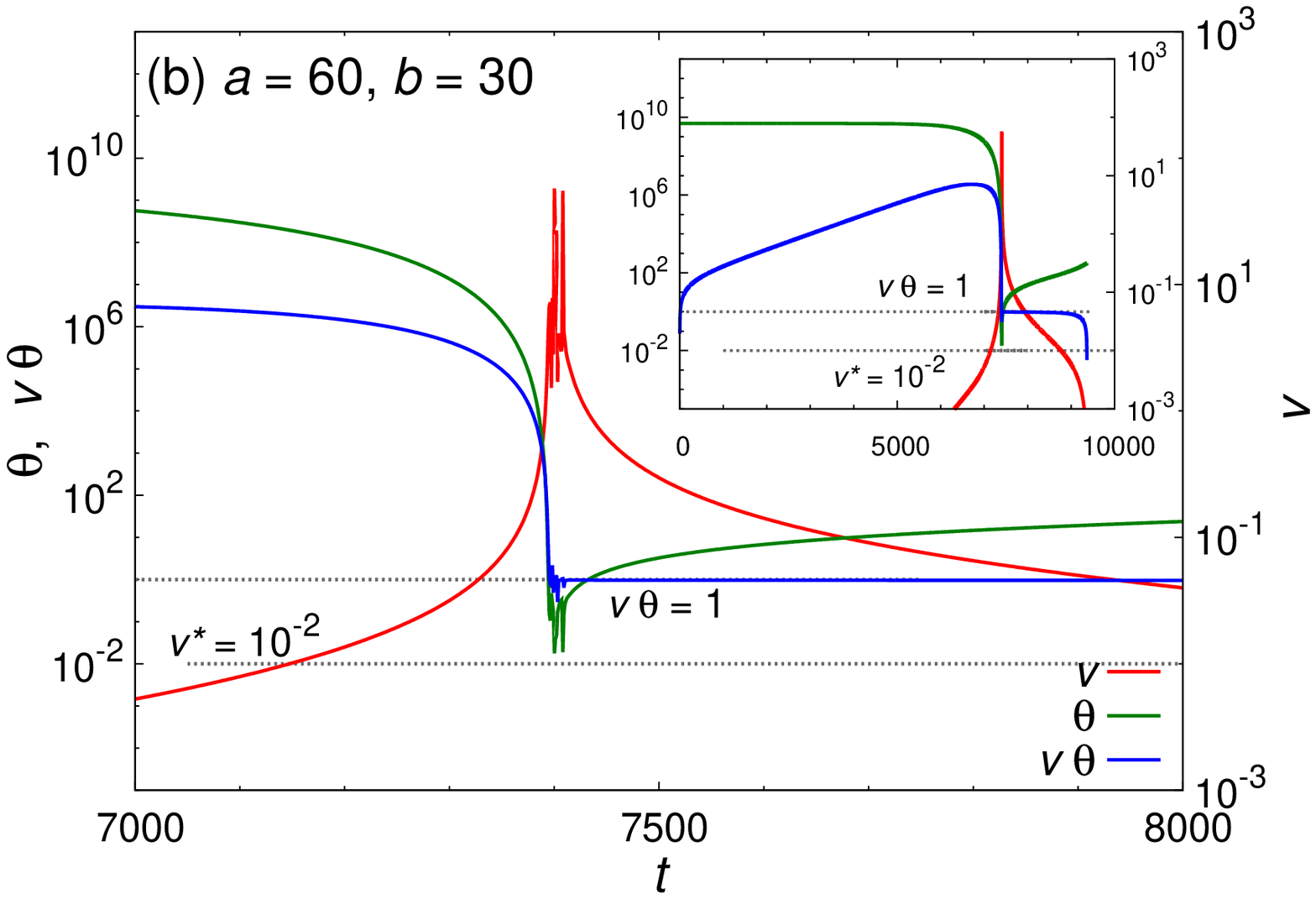}
\includegraphics[scale=0.45]{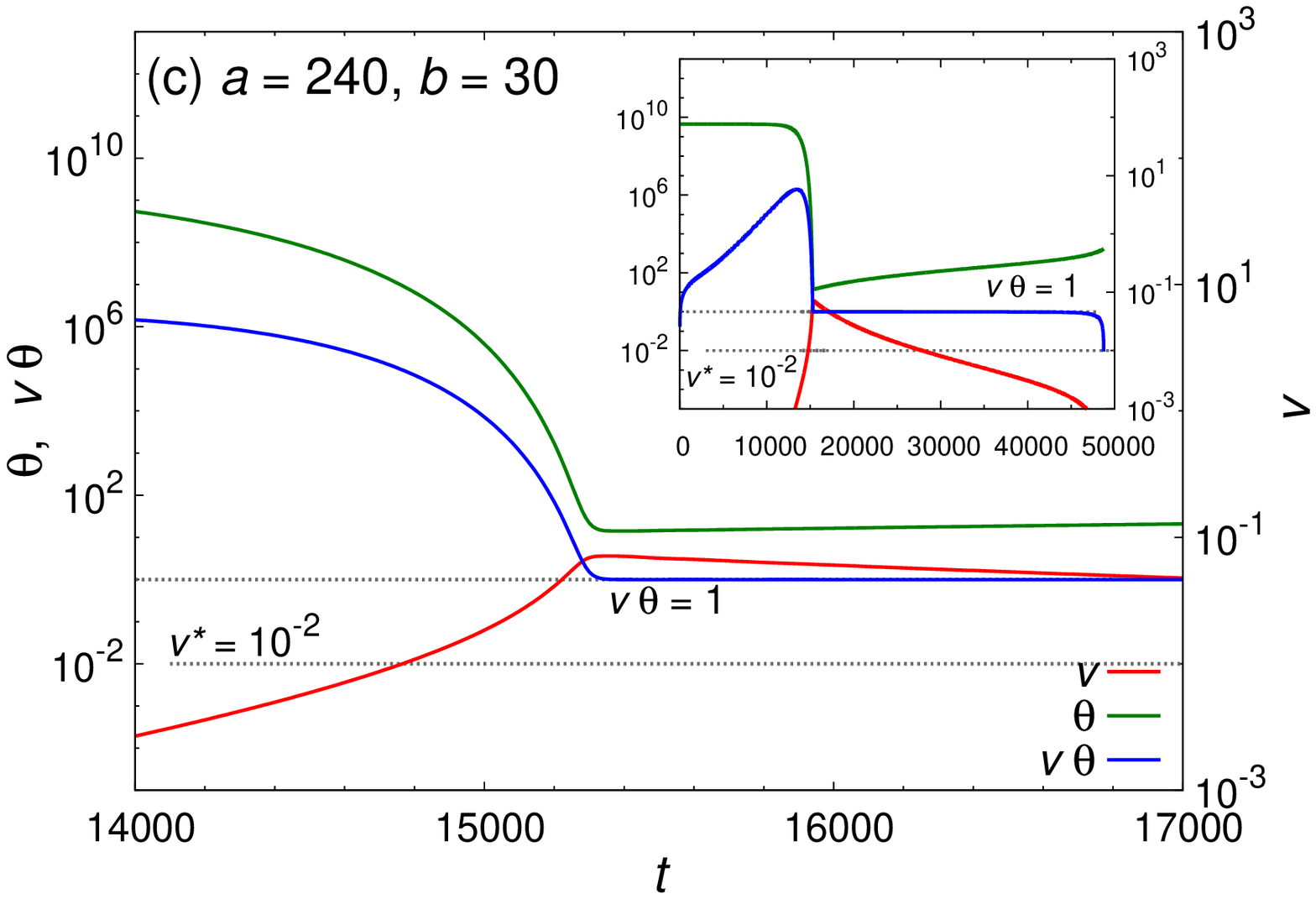}
\caption{
(Color online) The time dependence of the sliding velocity $v$ (right ordinate), the state variable $\theta$, and their multiple $v\theta$ (left ordinate) of an epicenter block for events in regions I, II and III. (a) $a=1$, (b) $a=60$, and (c) $a=240$, each corresponding to the events shown in Figs.1(a)-(c). Other parameters are $b=30$, $c=1000$, $v^*=10^{-2}$, $\ell=3$, and $\nu=10^{-7}$. The time origin $t=0$ is taken at the beginning of the nucleation process of the event. Insets show a longer time range, while the main panels are magnified views around the maximum $v$ region.
}
\end{figure}

 Slips with low sliding velocity are also realized in the nucleation process preceding main shocks \cite{Ueda2014,Ueda2015}. In fact, they are also realized even in silent earthquakes of region III. How such slow slips associated with the nucleation process resemble or differ from the slow slips studied here, and how they connect to the latter, are interesting questions. To gain insights into the issue, we show in Fig.4 the time dependence of several quantities for an epicenter block which experienced a slow nucleation process, including the sliding velocity $v$, the state variable $\theta$, and their multiple $v\theta$ for the events in regions I (a), II (b), and III (c), each corresponding to the events shown in Figs.1(a)-(c).

 As can be seen from Figs.4(b) and 4(c), the relation $v\theta=1$ is nearly satisfied in the slow-slip regimes. In fact, this relation implies an approximate stationarity, since $\theta$ is nearly constant from Eq.(3), and $v\simeq 1/\theta$. Thus, if $v$ takes a small value (or $\theta$ takes a large value) at the point where the relation $v\theta \simeq 1$ is met, long-lasting slow slips might follow, which actually seems to happen in the afterslip in region II, and in the silent earthquake in region III. Even in region I, the relation $v\theta=1$ is met, but only during the short period of high-speed rupture of a main shock around its maximum $v$.\cite{Ueda2015}

 The observation that the relation $v\theta \simeq 1$ is met during the slow slips of afterslips and silent earthquakes indicates that they are fundamentally different in nature from those of the precursory nucleation process where $v\theta >>1$, as can be seen from Fig.4. During the slow nucleation process, $\theta$ is of order $10^{10}$, representing a strongly stuck interface formed during the long interseismic period, and $v\theta$ is also large of order $10^{7}$, in spite of its low $v$-value. Then, as is evident from Eq.(3), $\theta$ actually decreases rapidly from its extremely large initial value. As such, the slow slip in the nucleation process is a far-from-stationary phenomenon in spite of its low sliding velocity. By contrast, at the slow slips realized in afterslips and silent earthquakes, the strongly stuck interface has been more or less released, with $\theta$ reduced to moderate values of order $10^0\sim 10^4$. The relation $v\theta \simeq 1$ is then well met, leading to near-stationarity. Another point to be noticed might be that the slow slip at silent earthquakes propagates in space with a constant rupture velocity, as can be seen from Fig.1(c), whereas the slow slip in the precursory nucleation process is spatially localized into the compact ``seed'' area. 

 Note that the relation $v\theta =1$ during the slow-slip process does not hold strictly, but only approximately. In fact, $v\theta $ is slightly smaller than unity throughout the slow-slip process, leading to a gradual increase of $\theta$ and an associated decrease of $v$. Finally, when $v$ reduces below $v^*$, the $a$ term changes its form due to the crossover effect, and the block motion comes to a complete stop, as shown in the insets of Figs.4(b) and (c). In other words, slow slip phenomena in afterslips and silent earthquakes are near-stationary, but are still essentially non-stationary phenomena.

 Finally, to examine the relevance of our present results to real seismic observations, we attempt order estimates of the slow-slip duration time $T$. Fig.2 indicates that the local duration time (the duration time for a given block) of a silent earthquake in region III might be $T\simeq 3\times 10^5$ for $v^*=10^{-3}$, which reads as several days in real time. Such a time scale may correspond to a short-term slow slip in real seismicity.\cite{HiroseObara,Ide} In fact, in our simulations, $T$ turns out to be inversely proportional to the largely unknown model parameter $v^*$ in the $v^*$-range studied ($10^{-1}\geq v^*\geq 10^{-3}$). If one extrapolates this to still smaller values of $v^*$, one gets $T\simeq 1$ year for $v^*=10^{-5}$, which may correspond to a long-term slow slip.\cite{Hirose,HiroseObara,Ide} 

  In summary, we studied the seismic properties of the 1D BK model obeying the RSF law over a broad parameter space, and we found a variety of slow-slip phenomena, including afterslips and silent earthquakes, which resembled those observed in real seismicity. We clarified the occurrence condition of such slow-slip phenomena, and the essential difference from the slow seismic slips as realized in the precursory nucleation process. Interestingly, the 1D BK model obeying the RSF law, being quite a simple model, is capable of reproducing rich seismic phenomena such as a main shock, a precursory nucleation process, an aftershock, and a silent earthquake by varying only a few model parameters.

 The authors are thankful to ISSP, the University of Tokyo for providing us with CPU time. This study is supported by a Grant-in-Aid for Scientific Research No.16K13851, and by the Ministry of Education, Culture, Sports, Science and Technology (MEXT) of Japan, under its Earthquake and Volcano Hazards Observation and Research Program.

\end{document}